\documentclass{article}
\usepackage{graphicx}
\usepackage[margin=1.0in]{geometry}

\usepackage{amsmath}
\usepackage{amsfonts}
\usepackage{ dsfont }
\usepackage{ dsfont }
\usepackage{ amssymb }
\usepackage{authblk}
\usepackage{braket}

\begin{document}

\title{Cardy's Conjecture and the Spectrum of Infrared Strongly-Coupled Quantum Field Theories}
\author[1]{Daniel Davies\thanks{Email: dadavies@ucsc.edu}}
\date{\today}
\affil[1]{\small{Department of Physics, University of California, Santa Cruz, Santa Cruz CA 95064}}

\maketitle

\begin{abstract}
Cardy's conjecture about the evolution of the trace anomaly under renormalization group (RG) flows is re-interpreted as an exact, non-perturbative statement about the scaling dimension of terms in the Lagrangian of the theory. When viewed in this way, the conjecture implies that field theories which are strongly coupled in the infrared (IR) may generically host states that are not manifest as solutions to the ultraviolet (UV) complete theory's equations of motion. In particular, the scaling dimension of operators in the Hamiltonian may deviate from their classical values by $\mathcal{O}(1)$ corrections in the IR, circumventing an old argument by Derrick about the non-existence of such states\cite{1}. We show that this framework provides a natural way to estimate the masses of these states using perturbation theory, and suggests a preferred reorganization of the degrees of freedom in the IR.

\end{abstract}

\newpage
\section{Introduction: Cardy's Conjecture}

Except in the rare cases(e.g. \cite{2}) of integrable or otherwise solvable models, it is generally impossible to derive by hand the spectrum of a quantum theory from only the Hamiltonian. Perturbation theory is therefore a standard technique for the professional theorist, but despite its many successes in describing phenomenological aspects of quantum theory, it is plagued by challenges\cite{3}\cite{4}. For the quantum field theorist, there are challenges aplenty; the first and most glaring issue are the short-distance divergences of Feynman diagrams when using the interaction picture to compute observables or scattering amplitudes. To amend this particular problem, one defines a `renormalized' version of their quantum theory, which introduces a mass scale $\mu$ that serves as an upper bound for the momentum-space resolution of the theory - or if one likes, a coarse-graining of spacetime. Since the theory is now manifestly insensitive to arbitrarily short distance events, the operators and parameters of the original theory must be modified in a $\mu$-dependent way to both avoid the divergences and make observable quantities independent of the choice of $\mu$. Since the parameters of the renormalized theory now depend on this scale (we say they ``run" with $\mu$), there is a potential problem with the application of perturbation theory should these parameters ever become large for a given value of $\mu$. This happens in Quantum Electrodynamics (QED): when renormalized at a very high scale the electric charge of the fermions become arbitrarily large. Since this limit is exactly where the renormalized theory meets the original QFT, such theories are generally considered to be unphysical, unless they are embedded into a new theory at some intermediate scale so as to avoid this issue (as is QED in Grand Unified Theories). On the contrary, we have theories like Quantum Chromodynamics (QCD) where the running of the coupling is opposite - the beta function is negative and the coupling becomes very large when renormalized at a low energy scale\cite{5}\cite{6}. QCD renormalized at scales much higher than a few hundred MeV is considered perturbative. However, perturbative QCD has failed to provide a description of the theory's spectrum which is consistent with observations of the long-distance physics\cite{7}. Given that there are many technical challenges with perturbation theory (Haag's theorem\cite{3}, non-convergence of power series expansions\cite{4}, general difficulties with calculating to high order) it should not be surprising that it generally fails to predict the spectrum of QFTs with negative beta functions.\\

Nevertheless, perturbation theory has taught us much about QFTs. The necessary introduction of the renormalized perturbation theory opened the door for the study of the Renormalization Group (RG)\cite{8}, from which we have learned much about non-perturbative physics. One such piece of knowledge is the c-theorem: in 2-dimensional conformal field theories (CFTs), there is a number $c[\mu]$ that enters as a proportionality constant in front of the anomalous divergence of the scale current. It has been established that c decreases monotonically\cite{9} as one flows to the IR ($\mu$ is decreased), regardless of the microscopic details of the theory. This establishes that (at least for 2 dimensional CFTs) a theory renormalized in the UV has more degrees of freedom than one renormalized in the IR. John Cardy conjectured that this is true for all field theories, and proposed a candidate for $c$ in higher dimensional, or non-conformal QFTs\cite{10}. A similar result has been proven in 4 dimensions, called the $a$-theorem\cite{11}. As with the c-theorem, $a$ is a multiplicative constant of the anomalous divergence of the scale current in a CFT. It is not the only anomalous term, but it is the component that survives integration over spacetime of the anomaly. In this context, one often sees the following ``equation" in the literature

\begin{equation}
\int d^D x \braket{T^\mu_\mu}\footnote{Here we are careful to mention that the use of $\mu$ is to indicate a spacetime index, not the scale at which the theory has been renormalized at. When we wish to make RG scale dependence explicit, we will use square brackets, e.g. $c[\mu]$} \sim \text{anomaly}
\end{equation}

In all relevant classical field theories it is the case that the divergence of the scale current is equal to the trace of the stress energy tensor. Cardy's conjecture, simply put, is that the left hand side of this equation is a monotonically increasing function of $\mu$. In two and four dimensions, the $c$ and $a$-theorems respectively, are examples of Cardy's conjecture for theories which exhibit a pattern of spontaneous conformal symmetry breaking. To endow an interacting field theory with conformal invariance, it is generally necessary to couple it to the metric tensor. At least in the case of the proof of the $a$-theorem, this is done through a massless mediator known as the dilaton. The dilaton-matter and dilaton-metric interactions are tuned so that the trace of the total stess-energy tensor vanishes identically. Additionally, the dilaton can be coupled arbitrarily weakly to the matter theory, allowing one to study the trace anomaly perturbatively as a consequence of the RG flow between conformal fixed points\cite{11}. This is possible to do because the IR effective theory of the dilaton and metric tensor is highly constrained by the assumed conformal (or at least Weyl) symmetry, and the anomaly coefficient appears in the four-derivative terms of the effective action. This is effectively how the $a$ and $c$ theorems were established. In three dimensions, any attempt to replicate the proof of the $a$-theorem is doomed, since there are no conformal(Weyl)-invariant terms constructed of the Riemann tensor or its derivatives with which to build an IR effective theory of the dilaton and metric \cite{12}. It is thus often claimed that there is no trace anomaly in three-dimensional quantum field theories. We wish to emphasize now that this is meant to be a statement about CFTs and their pattern of symmetry breaking in three dimensions, not for three-dimensional field theories in general. The present work does \textit{not} consider this pattern of symmetry breaking. We will focus on the physically relevant case of non-conformal theories, and whether conformal/scale symmetry is acquired in the UV is of no consequence to our results. Therefore, we do not need to assume any particular properties of the IR effective actions, and the methods and language used here may be quite different than most literature on Cardy's conjecture and CFTs. The use of the word `anomaly' to describe this phenomena then takes on a different meaning: instead of a violation of some classical conservation law (in general there is no conserved scale current), the trace anomaly serves as an obstruction to using the stress-energy tensor as a generator of scale transformations in the quantum theory.\\

The layout of this paper is as follows. In Section 2 we will make the ``equation" above more clear. In particular, we will argue that what belongs in those brackets (what has been called the divergence of the scale current) is not the trace of the stress-energy tensor, but rather a different tensor $\theta^\mu_\mu$ which only happens to equal $T^\mu_\mu$ when quantum mechanics is turned off. This argument is phrased entirely in the context of RG invariance and makes obvious what role the anomalous scale dimensions of operators should play in the discussion.\\

In section 3, once Cardys conjecture is translated into a statement about the anomalous scaling, we will briefly discuss a criteria for the existence of solutions to the field equations that makes direct use of these anomalous dimensions. It is noticed that for IR strongly coupled theories, the anomalous dimensions generically behave in such a way that allows solutions to become manifest in the field equations of the renormalized theory which are not present in the classical theory - something which the present author has conjectured might happen\cite{13}. The conditions on which such solutions become manifest are proposed.\\

\section{Scale Transformations and RG Invariance}
\subsection{Anomalous Dimension}
When a theory is defined by an arbitrary renormalization scale $\mu$, it no longer obeys the scaling relations expected from classical field theory. For example, consider the correlation function of renormalized fields $\hat{\phi}[\mu]$ with couplings $g_i[\mu]$
\begin{equation}
G^{(3)}(\mu;g_i, x_1,x_2,x_3) = \braket{\hat{\phi}(x_1)\hat{\phi}(x_2)\hat{\phi}(x_3)}
\end{equation}

\noindent It is useful to imagine the result of re-scaling these coordinates by a proportionality constant $\lambda$. From classical physics, we have the so-called `engineering dimension' $\Delta$ of the field defined by
\begin{equation}
\phi(\lambda x) = \lambda^{-\Delta}\phi(x)
\end{equation}

\noindent and naively, one would expect the dimension of $G^{(3)}$ to be $-3\Delta$. This is not the case, as a scale transformation of this form must be accompanied by a change of renormalization scale $\mu \rightarrow \mu^\prime(\lambda)$ as well. The correct result, which is consistent with demanding invariance under change of renormalization scale, is

\begin{equation}
G^{(3)}(\mu;g_i,\lambda x_1,\lambda x_2,\lambda x_3) = \lambda^{-3(\Delta + \gamma(g_i))}G^{(3)}(\mu^\prime;g_i, x_1,x_2,x_3)
\end{equation}

\noindent where $\gamma$ is the anomalous dimension of the field, and on the right hand side of this equation, it is understood that $g_i$ is being renormalized at the scale $\mu^\prime$. In a perturbative theory, $\gamma$ is usually a polynomial function in the couplings that starts at quadratic order. Thus it is small. However, the above equation should be interpereted as an exact statement about the relationship between scale transformations and RG flow, irrespective of the theorists ability to actually calculate such quantities. In a non-perturbative regime, the anomalous dimensions could be potentially $\mathcal{O}(1)$ or greater.\\

The uniqueness of the scale $\mu^\prime$ is also an issue. We suspect that as one increases $\lambda$ and probes the large distance properties of the UV renormalized theory, the corresponding choice of $\mu^\prime$ should decrease, equating to sensitivity of the IR theory. If $\mu^\prime$ is then a monotonically decreasing function of $\lambda$, we should also then suspect that the anomalous dimension does not take the same values twice at different scales, ensuring the uniqueness of the formula. These expectations, as we shall see, are at the heart of Cardy's conjecture and are tied to expectations that RG flows cannot be cyclic\cite{9}.\\

It is also imperative that we discuss the anomalous dimension of composite field operators, like $\hat{\phi}^2$. The anomalous dimension of this operator is not equal to twice that of the fundamental field $\hat{\phi}$. It is easy enough to check that the divergence structure of $\braket{\hat{\phi}^2(x)}$ and $\braket{\hat{\phi}(x)\hat{\phi}(y)}$ are different, and so their renormalized counterparts must subtract divergences differently and therefore scale differently. The exact scaling dimension of operators at separated points will still be only the sum of the scaling dimensions of each operator; this is a consequence of the cluster decomposition property\cite{15}. This rule however does not apply for composite operators in interacting field theories: one might consider this a consequence of the operator product expansion.\\

\subsection{Scale Currents}

In a classical field theory there is a straightforward, easy way to calculate the effects of a scale transformation on solutions to the field equations. One must only identify the right hand side of
\begin{equation}
\phi(x^\mu + \delta x^\mu) = \phi(x^\mu) + \delta\phi(x^\mu)
\end{equation}

\noindent which with the identification $\delta x^\mu = (\lambda-1)x^\mu$ is exactly the scale transformation mentioned previously. With $\lambda$ expanded in a power series around the value of 1, a scale current is obtained by variation of the action in the usual way prescribed by Noether's theorem. The result is

\begin{equation}
j_\mu = x_\nu \theta^{\mu\nu}
\end{equation}

\noindent for some symmetric tensor which is conserved ($\partial_\mu \theta^{\mu\nu} = 0$) if the action is manifestly translation invariant. In the case of gauge theories, it is possible to choose $\delta \phi$ ($\phi$ is meant as a stand-in for any field) in such a way that this quantity is gauge invariant. The divergence of this scale current is therefore the trace
\begin{equation}
\partial_\mu j^\mu = \theta^\mu_\mu
\end{equation}

In a classically scale-invariant theory, this vanishes. However even when the theory is not classically scale invariant, this is a useful quantity to know. By the equations of motion, the spacetime integral of $\theta^\mu_\mu$ can be found to be proportional to any mass terms that are implicit in the Lagrangian. The scale current is therefore a useful tool for analyzing the breaking of scale invariance(or a larger conformal invariance), be it spontaneous or explicit. Given a Lagrangian composed of ``operators" $\mathcal{O}_i(x)$ and their engineering dimensions $\Delta_i$:
\begin{equation}
\mathcal{L} = \sum_i \mathcal{O}_i(x)
\end{equation}

\noindent it is quite easy to compute the divergence of this current in $D=d+1$ dimensions. By definition of the engineering dimensions, it is
\begin{equation}
\theta^\mu_\mu(x) = \sum_i (\Delta_i-D)\mathcal{O}_i(x)
\end{equation}

In many reasonable classical field theories, it happens to be the case that $\theta^{\mu\nu} = T^{\mu\nu}$, the stress energy tensor. As in the case of gravitational physics, $T^{\mu\nu}$ can be computed by variation of the action with respect to the metric tensor. In General Relativity the metric tensor is a dynamical thing so $T^{\mu\nu}$ emerges in its equations of motion, but in non-gravitational field theories the variation with respect to the metric still computes the same quantity. For a static solution to the field equations whose stresses vanish at spatial infinity, the integral of $T^\mu_\mu$ is exactly the energy/rest-mass of the solution\cite{16}, in agreement with the interpretation of $\theta^\mu_\mu$. So long as there is no explicit coordinate dependence in the Lagrangian, these two tensors will be the same, classically.\\

All of this changes once quantum mechanics is turned on. Scale invariance (if it is present) is broken by the RG flow, and can only be effectively restored if the flow terminates at a fixed point\cite{17}. This can happen, but exact scale invariance is broken and $\braket{\hat{\theta}^\mu_\mu}$ is anomalous. It is possible that this `trace anomaly' is largely responsible for the existence of \textit{all} mass scales in nature. In particular, Non-Abelian Yang-Mills theory with no matter fields is classically scale invariant, and the excitations are massless. In the quantum theory however, a mass gap is conjectured\cite{7} and the theory dynamically acquires a scale which was not present classically.\\

Furthermore one should not expect that $\hat{\theta}^{\mu\nu} = \hat{T}^{\mu\nu}$ remains to be generically true once quantum mechanics is involved. One reason is quite simple: variation with respect to the metric is not a quantum mechanical operation. If the theory is not a quantum theory of gravity, there is no reason to suspect that the metric dependence of a quantum action is sensitive to the quantum nature of the matter fields. For example, if 
\begin{equation}
S_\text{int} = -\int d^D x \sqrt{-g}\hat{\phi}^4(x)
\end{equation}

\noindent then $\hat{T}^\mu_\mu$ necessarily contains a term $D \hat{\phi}^4(x)$. This is completely insensitive to the fact that the scaling dimension of this operator is dependent on what other operators are included in the Lagrangian as well as the scale $\mu$ that we have renormalized at. The derivation of $\hat{\theta}^\mu_\mu$ naturally leads to the inclusion of a term $-\left(\gamma_{\phi^4}-D\right)\hat{\phi}^4(x)$ which contains both information about the renromalization scale and the rest of the dynamics of the theory (here we have worked in a setting where $\Delta = 0$). If that is not enough to convince one that these tensors are different, consider the principle of RG invariance:

\subsection{RG Invariance and Cardy's Conjecture}

A principle tenant of the renormalization group procedure is the invariance of physical quantities with respect to $\mu$. The foremost example of a physical quantity, which is fundamental to the construction of any theory, is the invariant eigenvalue of the squared momentum operator $\hat{P}^2$. For a stationary state identified with a quanta of the theory, this is just the physical mass $M^2$. Consider a stationary state $\ket{\Psi}$, whose wavefunction vanishes sufficiently fast at spatial infinity. Manifest translation invariance implies, e.g. $\partial_\mu \hat{T}^{\mu i} =0$ or\footnote{This argument was borrowed from \cite{16}}
\begin{equation}
\int d^d x x_i \braket{\Psi|\partial_\mu \hat{T}^{\mu i}|\Psi} = \int d^d x x_i \partial_j\braket{\Psi| \hat{T}^{j i}|\Psi} = -\int d^d x \braket{\Psi|\hat{T}^i_i|\Psi} = 0
\end{equation}

Therefore, since the ``00" component of the stress energy tensor is the Hamiltonian, the spatial integral of $\braket{\Psi| \hat{T}^{\mu}_\mu|\Psi}$ is just the mass of the state, $M_\Psi$. For the vacuum, this is zero. How then could it be that the integrated trace of the stress energy tensor depends on $\mu$, for the vacuum or for other states? We argue of course that it does not, and that the operator relevant to Cardy's conjecture is not the true stress energy tensor, but rather $\hat{\theta}^\mu_\mu$ which generically differs from $\hat{T}^\mu_\mu$ by the $\mu$-dependent scale anomaly.\\

We will now re-state Cardy's conjecture. For a Lagrangian density as in eq. (8), where the terms are now true operators in a quantum theory, the divergence of the scale current is

\begin{equation}
\hat{\theta}^\mu_\mu(x) = \sum_i (\Delta_i + \gamma_i-D)\hat{\mathcal{O}}_i(x)
\end{equation}

\noindent where the RG scale dependence is now implicit in $\gamma_i$'s. Cardy's conjecture is then about the quantity 

\begin{equation}
 \Theta[\mu] \equiv\sum_i (\Delta_i + \gamma_i-D)\int d^D x \braket{\hat{\mathcal{O}}_i(x)} = \int d^D x\braket{\hat{T}^\mu_\mu} + \int d^D x \sum_i\gamma_i \braket{\hat{\mathcal{O}}_i}
\end{equation}

Specifically, $\Theta_\text{IR} \leq \Theta_\text{UV}$, i.e. that it is a monotonically increasing function of $\mu$. The first term in the right hand side is a constant: the anomaly is represented as the remainder. Of course there are two dependencies on $\mu$ in the anomaly: that of $\gamma_i$'s and that of the expectation values $\braket{\hat{\mathcal{O}}_i(x)}$. We claim that a model-independent interpretation of Cardy's conjecture is really just a statement about the anomalous dimensions: $\gamma_i[\mu]$ increases monotonically as we flow to the IR. This would in fact affirm our earlier assumption that eq. (4) necessitated a unique choice of $\mu^\prime$. As a reminder, were this not the case we should conclude that RG flows can be cyclic, something which certainly should not describe any physical system. For the remainder of this paper we will work under the assumption that this is the correct interpretation of Cardy's conjecture, and that the conjecture is correct.\\

\section{Finding New States}

Briefly, let's take a detour back to the land of classical field theory. Consider the case of an interacting Klein-Gordon field in 2+1 dimensions, with a $\mathbb{Z}_2$ potential in its spontaneously broken phase (perhaps $V(\phi) = -m^2\phi^2 + g_6 \phi^6 + V_0$). Naively, one might suspect that domain walls are formed, and could be stable. This is not the case. We divide the Hamiltonian into kinetic and potential terms

\begin{equation}
H = K+U
\end{equation}

\noindent where K and U are positive definite. We now imagine scaling a static domain wall of spatial size 1 to a size $\lambda^{-1}$. Since $\Delta_K = 2$ and $\Delta_U = 0$ we have in $d=2$ spatial dimensions:

\begin{equation}
H[\lambda] = K + \lambda^{-2}U
\end{equation}

\noindent which cannot be minimized except at $\lambda = \infty$. Therefore the domain wall is unstable to shrinking indefinitely (this is of course due to the unbalanced tension on the wall). Furthermore, no such extended field configurations exist and are stable in this model. The above classical scaling argument is the core of what is known as Derrick's theorem\cite{1}.\\

\subsection{Large Anomalous Dimensions}

The story is potentially much different in the quantum theory. It is easy to see how different things can be by assuming at first $K$ and $U$ acquire anomalous dimensions $\gamma_K$ and $\gamma_U$. The conditions for a stable state then become the condition for a local minima of $H[\lambda]$ at $\lambda = 1$:
\begin{align}
\gamma_K K + (\gamma_U - 2)U &= 0\\
\gamma_K (\gamma_K -1) K + (\gamma_U - 2)(\gamma_U - 3)U &> 0
\end{align}

Alternatively (by eliminating $U$ for $K$ using eq. (16)), $(2-\gamma_U)/\gamma_K> 0$ and $\gamma_K(2+\gamma_K +\gamma_U) > 0$. Now obviously in a real model the terms that appear in the potential energy will not all have identical scaling dimensions. The argument is clear though; the activation of anomalous scaling of operators that contribute to the energy density of a state can circumvent the conclusions of Derrick's theorem. This of course depends on the relative signs and magnitudes of the anomalous dimensions in question. This is where Cardy's conjecture enters the picture.\\

If the theory is weakly coupled in the IR, $\gamma_K$ and $\gamma_U$ are small, close to zero. As one increases the RG scale, these terms either decrease or stay the same. At some UV scale then, they stand the chance of being large (perhaps before the theory is embedded to become UV complete) but negative. The stated conditions for circumvention of Derrick's theorem are then not satisfied.\\

This theory in 2+1 dimensions happens to have a negative beta function - it is strongly coupled in the IR and weakly coupled in the UV\cite{18}. Suppose then that in the UV where the anomalous dimensions are close to zero (and possibly negative) Derrick's theorem is not circumvented. Cardy's conjecture implies then that as we flow to the IR, $\gamma_K$ and $\gamma_U$ grow to become large as we approach the non-perturbative regime. If the flow does not terminate at a fixed point before this happens, the anomalous dimensions stand a chance of approaching unity (and positive), where the above conditions can almost certainly be satisfied. The meaning of the title of this paper is now revealed: it is precisely the case of an IR strongly-coupled QFT where one expects new massive solitonic states to become manifest in the field equations. That they are not manifest as solutions along the entire RG trajectory should not be surprising: perturbation theory in the UV is by construction only sensitive to a subsection of the total Hilbert space of a QFT\cite{13}.\\

We did not just prove that stable domain walls will exist in the 2+1 dimensional scalar field theory described at the beginning of this section. Rather, we used this simple case to visualize the potential for strongly-coupled theories to host solutions which have no classical analogue. Whether a theory actually hosts such solitonic states must be checked on a case-by-case level, and is subject to details that we will not address here. In upcoming works we will investigate specific cases and demonstrate that signifigant deviations from classical estimates of soliton masses may occur.\\

\subsection{The Masses}

Suppose a theorist makes some exact non-perturbative evaluation of the operator dimensions in the Hamiltonian at some UV scale $\mu$, and uses the beta function for all relevant coupling constants to put bounds on how slowly those dimensions evolve with $\mu$. This theorist then discovers that by our monotinicity arguments, conditions (16) and (17) are met as one flows to the IR, at an RG scale no smaller than $\mu^*$. What then is the mass of the state which emerges there, and how does it depend on the value of $\mu^*$?\\

We don't have an exact answer to this question of course, and necessarily this is a regime where perturbative calculations of anomalous dimensions are likely not very accurate. But if the conditions above can be met, one should be able to put a lower bound on the mass of the lightest solitonic state. Generically we should consider a time-independent state (i.e. a static field configuration) and a Hamiltonian of the form

\begin{align}
H = \sum_i H_i[\mu] + \Omega[\mu]
\end{align}

\noindent where the $H_i$ are derived from the integrated expectation values (with respect to the massive state, not the vacuum) of operators in eq. (8), and $\Omega$ is a free energy. Often one will encounter texts which claim this free energy is a cosmological constant, but this is not correct: while renormalization of vacuum graphs requires the introduction of a field-independent term in the action to regulate divergences, this is an RG-variant term and is compatible with a zero vacuum energy. The true meaning of $\Omega[\mu]$ is that of the free energy associated with the invisible fluctuation modes with momenta greater than $\mu$. Both $\Omega$ and the $H_i$ are RG-variant quantities, but by construction their sum should not be.\\

What matters now is essentially applying stability conditions (16) and (17) to our Hamiltonian (18). The result is not literally (16) and (17), but rather (in $d$ spatial dimensions)

\begin{align}
&\sum_i (\Delta_i+\gamma_i - d)H_i + (\gamma_\Omega -d)\Omega = 0\\
&\sum_i (\Delta_i+\gamma_i - d)(\Delta_i+\gamma_i - d+1)H_i + (\gamma_\Omega -d)(\gamma_\Omega -d+1)\Omega  > 0
\end{align}

Here $\Omega$ is assumed to have no engineering dimension, and only acquires dimension $\gamma_\Omega$ through the anomalous running of the renormalized coupling constants of which it is a function. Assuming that conditions (19) and (20) are met at and below the RG scale $\mu^*$, we now have known relationships between the numerical values of $H_i[\mu^*]$ and $\Omega[\mu^*]$. As is typical with classical extended field configurations, all of these terms should generically differ only by $\mathcal{O}(1)$ coefficients\cite{19}. The anomalous dimensions will generically\footnote{The previous 2+1 dimensional example was a special case. There it seems that these solutions turn on as soon as $\gamma$'s are positive, even if very small. However, in this number of dimensions many models already host solitonic states, e.g. vortices, a consequence of the fact that $\Delta_K = d = 2$. In higher dimensions, where soliton solutions are generically not present, the $\gamma$'s have more work to do and therefore will be $\mathcal{O}(d-\Delta)$ at $\mu^*$} be $\mathcal{O}(d-\Delta)\sim\mathcal{O}(1)$ at $\mu^*$ and since the coefficients of the $H_i$'s and $\Omega$ in eqs. (19) and (20) will be differing in sign, we argue that $H \sim \Omega[\mu^*]$ is as good an estimate for the lower bound of the mass as any. What is the value of $\Omega[\mu^*]$? Naively, it is some integral of the free energy density $\omega(g_i)$, a function of the couplings of the theory renormalized at $\mu^*$. The correct expression should depend on some normalized energy density profile of the state, and since this is at the moment indeterminate, we conjecture the following:

\begin{align}
M \sim \Omega[\mu^*] \gtrsim \frac{\omega[\mu^*]}{(\mu^*)^d}
\end{align}

Since the state only manifests as a solution to our quantum equations of motion when we are insensitive to distance scales less than $1/\mu^*$, the volume of the physical object should be at least $(1/\mu^*)^d$. The natural guess is then eq. (21).

\section{Conclusion and Going Forward}

We have interpreted Cardy's conjecture as a statement about the change of anomalous dimensions under RG flow. The basic requirements that RG flows not be cyclic and that an IR renormalized theory have less degrees of freedom than a UV renormalized theory are realized if some scaling dimensions are larger in the IR than in the UV. This becomes consequential if those dimensions deviate from classical values by $\mathcal{O}(1)$ corrections, and those deviations are positive. This happens when the theory is IR strongly coupled, and it opens the door for a circumvention of Derrick's classical no-go theorem. Should new solutions to the renormalized field equations emerge, we identify them as solitons and have proposed a way to estimate a lower bound on their mass using perturbation theory in the UV. Such a thing is only possible because of monotinicity arguments. Once the solutions are found and characterized, we propose a reorganization of the degrees of freedom at scales at and below $\mu^*$ to reflect the manifestation of these new solutions. One should expand perturbatively around such solutions, rather than the free field theory configurations.

\end{document}